\documentclass[aps,prd,nofootinbib,preprint,floatfix,unsortedaddress]{revtex4-1}
\usepackage{setspace}
\usepackage{graphicx}
\usepackage{dcolumn}
\usepackage{multirow}
\usepackage{subfigure}
\usepackage{times,mathptm}
\usepackage{float}
\usepackage{color}
\usepackage{amsmath,amsfonts}
\usepackage{mathptmx}
\usepackage{mathrsfs}
\usepackage{bbm}
\usepackage{bm}
\usepackage{xfrac}

\newcommand{\bea}{\begin{eqnarray}}
\newcommand{\eea}{\end{eqnarray}}

\renewcommand{\d}{\delta}

\renewcommand{\l}{\lambda}

\newcommand{\tQ}{\widetilde{Q}}

\renewcommand{\b}{\beta}
\renewcommand{\a}{\alpha}

\newcommand{\tr}{\text{Tr}}

\newcommand{\vx}{{\vec{x}}}
\newcommand{\vy}{{\vec{y}}}

\newcommand{\vk}{{\vec{k}}}

\newcommand{\m}{\mu}
\newcommand{\g}{\gamma}

\newcommand{\s}{\sigma}
\renewcommand{\k}{\kappa}

\newcommand{\D}{\Delta}

\newcommand{\oh}{\frac{1}{2}}

\newcommand{\dg}{\dagger}
\newcommand{\non}{\nonumber}

\newcommand{\rf}[1]{(\ref{#1})}
\newcommand{\ra}{\rightarrow}
\newcommand{\pa}{\partial}
\renewcommand{\vec}[1]{\bm #1}

\usepackage{ulem}

\begin{document}

\bibliographystyle{h-physrev5}

\title{Effective Polyakov line action from the relative weights method} 
 
\author{Jeff Greensite}
\affiliation{Niels Bohr International Academy, Blegdamsvej 17, DK-2100
Copenhagen \O, Denmark}
\altaffiliation[Permanent address: ]{Physics and Astronomy Dept., San Francisco State
University, San Francisco, CA~94132, USA}

\author{Kurt Langfeld}
\affiliation{\singlespacing School of Computing \& Mathematics, University of Plymouth, Plymouth, PL4 8AA, UK}
\date{\today}
\vspace{60pt}
\begin{abstract}

\singlespacing
 
      We apply the relative weights method (arXiv:1209.5697) to determine the effective Polyakov line action for SU(2) lattice gauge theory in the confined phase, at lattice coupling $\b=2.2$ and $N_t=4$ lattice spacings in the time direction.  The effective action turns out to be bilinear in the fundamental representation Polyakov line variables, with a rather simple expression for the finite range kernel.  The validity of this action is tested by computing Polyakov line correlators, via Monte Carlo simulation, in both the effective action and the underlying lattice theory.  It is found that the correlators in each theory are in very close agreement.    
\end{abstract}

\maketitle

\singlespacing
\section{\label{sec:intro}Introduction}
 
     The Polyakov line action (PLA) is an action obtained from lattice gauge theory
 when all degrees of freedom are integrated out, under the constraint that the Polyakov line holonomies are held fixed.  There are some indications  \cite{Gattringer:2011gq,*Mercado:2012ue,Fromm:2011qi,Aarts:2011zn,Greensite:2012xv} that the sign problem in this theory, at non-zero chemical potential, may be more tractable than the sign problem in the underlying lattice gauge theory (for a review, cf.\ \cite{deForcrand:2010ys}), and if so it could provide us with a new tool for investigating the QCD phase diagram.  It is fairly straightforward, given the PLA at chemical potential $\m=0$, to introduce a non-zero chemical potential, as we discuss in section \ref{sec:conclude}.   The problem we address here is how to extract the PLA from the underlying lattice gauge theory at $\m=0$. 
  
    This article is a follow-up to ref.\ \cite{Greensite:2012dy}, which presented a novel ``relative weights" technique for deriving the PLA, based on a method used previously in studies of the Yang-Mills vacuum wavefunctional \cite{Greensite:2011pj,*Greensite:1988rr}.  The method was tested at strong couplings, where the answer is known, and a conjecture for the action at a weaker coupling, for SU(2) pure gauge theory at $\b=2.2$ and inverse temperature $N_t=4$ lattice spacings, was presented.  This conjecture was based, however, on a study limited to fairly atypical regions of field configuration space.  Below we will apply the method in the region expected to dominate the path integral, and a rather different (and in fact simpler) action from the one conjectured in ref.\ \cite{Greensite:2012dy} emerges.  As a crucial test of the derived PLA, we compute the two-point Polyakov line correlator from Monte Carlo simulations of both the PLA and the underlying gauge theory.  These correlators will be seen to agree quite accurately.  Our effective PLA turns out to be only bilinear in the Polyakov line variables, with a simple expression for
the finite range kernel.
    
    There have been a number of previous attempts to derive the PLA from lattice gauge theory at finite temperature.
These include strong-coupling expansions \cite{Fromm:2011qi}, the Inverse Monte Carlo method \cite{Dittmann:2003qt,*Heinzl:2005xv}, and the demon approach \cite{Velytsky:2008bh,*Wozar:2008nv}.  All of these methods generate effective Polyakov line actions of varying degrees of complexity.  We believe, however, that an accurate agreement of Polyakov line correlators in the confined phase, computed in the effective and underlying lattice gauge theories, has not been demonstrated in any of the previous studies, at least not beyond two or three lattice spacings in Polyakov line separation.  

   It should also be mentioned that there are a number of studies which are concerned with deducing the Polyakov line potential, with particular application to the deconfinement transition, c.f.\ \cite{Dumitru:2012fw} and references therein.  There have also been efforts, e.g.\ \cite{Langfeld:1999ig}, to express the fermion determinant in terms of a potential involving Polyakov lines.  These studies do not arrive at a full PLA as defined above, and hence their focus is somewhat different from ours.        
        
    Our article is organized as follows:  An improved version of the relative weights method is presented in section \ref{sec:rw} below.  The technique is applied to pure SU(2) gauge theory in section \ref{sec:grad}, again at $\b=2.2$ and $N_t=4$, and the Polyakov line correlators of the derived PLA are compared to those of lattice gauge theory.  Application to a gauge-Higgs theory, with a scalar matter field explicitly breaking global $Z_2$ center symmetry, is presented in section \ref{sec:gh}.  Section 
\ref{sec:conclude} contains our conclusions.  The extension of our method to gauge theories with dynamical fermions
is discussed in an appendix.

\section{\label{sec:rw}The relative weights method}

     The relative weights method, as applied to deriving the PLA $S_P$, was introduced in ref.\ \cite{Greensite:2012dy}.  The technique is particularly well adapted to computing path (or ``directional'') derivatives of the effective action $S_P$ in the space of all Polyakov line configurations.  A single configuration $\{U_{\vx}\}$ is a point in this space, where we specify the group-valued Polyakov line holonomies $U_{\vx}$ at each spatial point $\vx$ in a three-dimensional volume.  Let $\{U_\vx(\l)\}$ be a path through this space of configurations, where $\l$ parametrizes the path.  The relative weights method computes the path derivative $\pa S_P[U_\vx(\l)] /\pa \l$ at some given $\l=\l_0$.  In this section we present a variant of the relative weights approach which, while equivalent to the original method of \cite{Greensite:2012dy}, is numerically more efficient. 
     
     In order to minimize minus signs later on, we adopt the convention that the Boltzmann weight is proportional to
$\exp[+S_P]$.  Let $S_L$ be the lattice gauge action on an $L^3 \times N_t$ volume with coupling $\b$ for the Wilson action.  
$S_L$ may contain pseudofermion or bosonic matter degrees of freedom, collectively denoted by $\phi$.  It is convenient to go to temporal gauge, so that all timelike link variables are set to the unit matrix except on a timeslice at $t=0$.  Then the PLA $S_P$ is defined as
\bea
\exp\Bigl[S_P[U_{\vx}]\Bigl] =    \int  DU_0(\vx,0) DU_k  D\phi ~ \left\{\prod_{\vx} \d[U_{\vx}-U_0(\vx,0)]  \right\}
 e^{S_L} \ .
\label{S_P}
\eea
Because of the residual $U_0(\vx,0) \ra g(\vx) U_0(\vx,0) g^\dg(\vx)$ symmetry in temporal gauge, it follows that $S_P$ can only
depend on the eigenvalues of the $U_{\vx}$ matrices.

    While the functional integration in \rf{S_P} can only be carried out in special cases, e.g. via strong coupling and hopping parameter expansions valid a certain range of parameters, the ratio (or ``relative weights'') $\exp[S_P[U'_{\vx}]]/\exp[S_P[U''_{\vx}]]$ evaluated at nearby configurations $U'_{\vx},~U''_{\vx}$ is calculable numerically.  This fact enables us to compute
path derivatives of $S_P$.  

    Let us consider a set of $M$ Polyakov line configurations 
\bea
 \Bigl\{ \{U^{(n)}_{\vx}, \mbox{all~} \vx \}, ~ n=1,2,...,M \Bigr\} \ ,
\label{set}
\eea 
corresponding to values of the path parameter
\bea
           \l_n = \l_0 + \left( n - {M+1 \over 2}\right) \D \l ~~~,~~~ n=1,2,...,M ~~~  \ ,
\eea
and define
\bea
    S_L^{(m)}[U,\phi] \equiv  S_L\Bigl[U_0(\vx,0)=U^{(m)}_\vx,U_k(x,t),\phi(x,t)\Bigr]
\eea
to be the lattice action in temporal gauge with the timelike links at $t=0$ fixed to the $m$-th member of the set \rf{set}.  We also
define 
\bea
    \D S_P^{(m+1)} &\equiv& S_P[U^{(m+1)}] - S_P[U^{(m)}]
\non \\
        Z_m &\equiv& \int DU_k D\phi ~ e^{S_L^{(m)}} \ .
\eea
From \rf{S_P}, we have
\bea
     \exp[\D S_P^{(m+1)}] &=& {\exp\Bigl[S_P[U^{(m+1)}\Bigr] \over \exp\Bigl[S_P[U^{(m)}]\Bigr] } 
\non \\
    &=&  {\int DU_k D\phi  ~ e^{S_L^{(m+1)}} \over \int DU_k D\phi ~ e^{S_L^{(m)}}  }                               
\non \\
   &=&    {\int DU_k D\phi  ~ \exp\Bigl[S_L^{(m+1)}-S_L^{(m)}\Bigr] e^{S_L^{(m)}}\over 
               \int DU_k D\phi  ~ e^{S_L^{(m)}}  }
\non \\
   &=&   \left\langle   \exp\Bigl[\D S^{(m+1)}\Bigr] \right\rangle_m \ ,
\eea
 where  $\langle ... \rangle_m$ indicates that the expectation value is taken from ensembles with Boltzmann factor
 $\exp[S^{(m)}]/Z_m$.    For sufficiently small $\D \l$,
\bea
        {d S_P[U_\vx(\l)] \over d\l}  &\approx&    {\D S_P^{(m+1)} \over \D \l}
\non \\
                &=& {1\over \D \l}  \log\left( \left\langle   \exp\Bigl[\D S^{(m+1)}\Bigr] \right\rangle_m  \right)   \ ,                                          
\eea
and these should closely agree, for all $m<M$, with the derivative evaluated at the central value of $\l=\l_0$.  We can then improve our estimate by making use of all $M$ configurations, taking the average of derivatives
\bea
\left({d S_P[U_\vx(\l)] \over d\l}\right)_{\l=\l_0} \approx {1\over \D \l} {1\over M-1} \sum_{m=1}^{M-1}
 \log\left( \left\langle \exp\Bigl[\D S^{(m+1)}\Bigr] \right\rangle_m \right) \ .
\label{snake}
\eea
 
    The question then becomes which point $\{U_{\vx}(\l_0)\}$ in configuration space should be chosen for the computation, and which directional derivatives $dS_P/d\l$ at this point should computed, in order to deduce $S_P$.   It is possible that the
choice is not very important, and that $S_P$ is well approximated by the same simple expression everywhere in configuration
space.  However, if this is not the case, then calculating path derivatives in some very atypical corner of configuration space
may lead to an approximate answer for $S_P$ which may be correct in that particular corner, but misleading in the bulk of configuration space.  

   In ref.\ \cite{Greensite:2012dy} the path derivatives were computed using three types of sets \rf{set} for SU(2) lattice gauge theory.  These were (i) Polyakov lines which were constant in space, and the $\l$ parameter was the amplitude of 
$P_{\vx}= \oh \tr[U_{\vx}]=\l$; (ii) Polyakov lines which consisted of small plane wave fluctuations around a constant background, $P_{\vx} = P_0 + \l \cos(\vk \cdot \vx)$ with $\l \ll P_0$; and (iii) Polyakov lines in which $P_{\vx}$ varied 
as $P_{\vx}= \l \cos(\vk \cdot \vx)$.  Compared to thermalized
timeslice configurations $U_0(\vx,0)$ generated in a normal lattice Monte Carlo simulation, such configurations are very
atypical.  In a thermalized configuration in the confined phase, the Fourier components of the configuration are all of
$O(1/\sqrt{V_3})$, where $V_3=L^3$ is the lattice volume of the D=3 dimensional timeslice, whereas in the special configurations just mentioned, one Fourier component (which may be the zero mode) is of $O(1)$.   For computing the PLA at a strong lattice coupling, where this action can be evaluated via a strong coupling expansion, the atypical nature of the constant + plane wave, or pure plane wave configurations did not seem important, and the PLA deduced from the relative weights data was a close match to the known result.   There was no similar result to compare to at $\b=2.2, ~ N_t=4$, and although an expression for $S_P$ matching the results for $dS_P/d\l$ was deduced from fitting the data, there is a concern that this expression might only be valid in the special region of configuration space where it was derived.

     We will investigate the action in more typical regions of configuration space in the next section.

\section{\label{sec:grad}Derivatives of $\mathbf{S_P}$ in a thermalized background}

   The goal is to use the expression for the path derivative \rf{snake} to determine $S_P$.  However, if $S_P$ does not have a simple form everywhere in configuration space (and it may not), then it is at least required that we have a fairly accurate approximation to $S_P$ in the region which is important for the computation of observables, i.e.\ the region occupied by typical thermalized configurations  $\{U_{\vx}\}$.   A set of timelike link configurations $\{U_0(\vx,0)\}$
on the $t=0$ timeslice,  generated by a numerical simulation of the underlying lattice gauge theory, is a sample of
such configurations.  Let us define, for the SU(2)
gauge group that we will consider here,
\bea
                P_{\vx} \equiv \oh \tr[U_{\vx}] =  \oh \sum_{\vk} \Bigl\{ a_{\vk} \cos(\vk \cdot \vx) + 
                       b_{\vk} \sin(\vk \cdot \vx) \Bigr\} \ ,
\label{dft}
\eea
where the sum  runs over all wavevectors $\vk$ on a cubic lattice of volume $L^3$, and ${a_{\vk}=a_{-\vk}, ~ b_{\vk}=-b_{-\vk}}$    are real-valued. Then we may consider calculating numerically, by the relative weights method, derivatives with respect to
the Fourier components
\bea
             \left( {\pa S_P \over \pa a_{\vk}} \right)_{a_{\vk}=\a}  ~~,~~   \left( {\pa S_P \over \pa b_{\vk}} \right)_{b_{\vk}=\a}  \ ,
\label{grad1}
\eea
in a background in which all other Fourier components are drawn from a thermalized configuration. By calculating the
derivative for some range of $\a$, it is possible to extrapolate
to small $\a$ of order $1/\sqrt{V_3}$, which is the typical magnitude of a Fourier component in thermalized configurations. It is sufficient in practice to concentrate on the coefficients of the cosine terms in the Fourier expansion, since the sine terms give similar results.  We then try to reconstruct $S_P$ from this information.

   There are potentially two obstacles to this approach.  First, there are as many independent Fourier components as there are lattice sites in the $V_3$ volume, and this is too many to calculate in practice.  Secondly, it might be that the results are strongly dependent on the particular thermalized background which is used.   Concerning the first obstacle, this will not be a problem if the derivatives with respect to $a_{\vk}$ have a simple dependence on the lattice momentum
\bea
            k_L = \sqrt{4 \sum_{i=1}^3 \sin^2(\oh k_i)} \ ,
\label{kL}
\eea
which can be deduced from a small sample of all possible components.
As for the second obstacle, this is not a problem if it turns out that the dependence of the final results on the particular choice of thermalized configuration is very weak.

\subsection{Deriving the effective PLA}

      The first step in the extrapolation to small $\a$ is to run a standard lattice Monte Carlo, stop at some thermalized configuration, and calculate all the Polyakov line holonomies,
\bea
          {\cal P}_\vx &=& U_0(\vx,1) U_0(\vx,2)...U_0(\vx,N_t)
\non \\
     &=& d_4(n_1,n_2,n_3) \mathbbm{1} + i \vec{d}(n_1,n_2,n_3) \cdot \vec{\s}  \ ,
\label{tslice}
\eea
where $\vx=(n_1,n_2,n_3)$, in that configuration.   Define $W(\vx)=d_4(\vx)$.
We pick a particular wavevector $\vk$ which is specified by three integers 
$(m_1,m_2,m_3)$ with corresponding wavenumber components
\bea
          k_i = {2\pi \over L} m_i \ ,
\label{waveno}
\eea 
and set the coefficient of $\cos(\vk \cdot \vx)$, in the sine-cosine expansion of $W(\vx)$, to zero.  Denote
the modified array, with the $\cos(\vk \cdot \vx)$ term removed,  as $W'(\vx)$.

    Next, construct a set of $M=20$ configurations with:
\bea
            P^{(n)}_\vx &=&  a_\vk^{(n)} \cos(k_1 n_1 + k_2 n_2 + k_3 n_3) + (1-\a-\d)W'(\vx)
\non \\
            a_\vk^{(n)} &=& \a + \Bigl( n - \oh (M+1)\Bigr) \g /L^3 ~~~,~~~ n=1,2,...,M \ ,
\label{config1}
\eea
where $\g=L^3 \D a$ is a constant chosen to be as small as possible, but still large enough to get some
spread in the data. Typically $\g \approx 0.5$.  

   The factor $1-\a-\d$ in \rf{config1} is introduced in order to keep $P^{(n)}_\vx$, with rare exceptions, inside the range $[-1:1]$. 
Ideally one would like to leave all $\{a_{\vk'},b_{\vk'}\}$ in the thermalized configuration unaltered, apart from the mode with
$\vk'=\vk$, i.e. $P^{(n)}_\vx = a_{\vk} \cos{\vk \cdot \vx} + W'(\vx)$.   At finite $\a$, however, this has the disadvantage that at
many sites $|P_\vx| > 1$.   To see this, note that  $W(\vx)$, from which $W'(\vx)$ is derived, may come close to the limits 
$\pm 1$ at some sites, and at these sites the additional contribution $a_{\vk} \cos(\vk \cdot \vx)$ may put the sum outside the allowed range by as much as $\a$.  Moreover, by removing the $\cos(\vk \cdot \vx)$ mode, $W'(\vx)$ may already lie outside the range $[-1:1]$ at some sites; this is especially true for $\vk=0$. We must also allow for the fact that half of the $\{a_\vk^{(n)}\}$ are slightly greater than $\a$.  For this reason we reduce the amplitude of the added thermalized configuration by a factor of $1-\a-\d$ (in our simulations we found $\d=.04$ sufficient).  In the exceptional cases where 
$P^{(n)}_\vx$ still lies outside the allowed range, it is truncated to the nearest limit, i.e. $\pm 1$.

   We then construct the SU(2) variables, at each site, which have $2 P^{(n)}_\vx$ as the trace
\bea
         U^{(n)}_\vx =  P^{(n)}_\vx  \mathbbm{1} + i  s^{(n)}\vec{d}(\vx) \cdot \vec{\s}        \ , 
\label{config2}
\eea
where, to insure unitarity,
\bea
         s^{(n)} = \sqrt{1 - (P^{(n)}_\vx)^2 \over \vec{d}(\vx) \cdot \vec{d}(\vx)} \ .
\label{config3}
\eea 
     The calculation of $(\pa S_P/\pa a_{\vk})_\a$ proceeds as described above.  For the choice of
$U^{(n)}$ given above, it is easy to see that for a lattice of extension $L$ in the spatial directions
\bea
{1\over L^3} \left({\pa S_P[U_\vx(a_{\vk})] \over \pa a_{\vk}}\right)_{a_{\vk}=\a} \approx  {1\over \g (M-1)} \sum_{m=1}^{M-1}
 \log\left( \left\langle \exp\Bigl[\D S^{(m+1)}[U]\Bigr] \right\rangle_m \right) \ .
\eea
The results for the derivatives are found to depend only weakly on the choice of thermalized time slice \rf{tslice} generated in an ordinary Monte Carlo run.  The dependence is most pronounced at small $k_L$, with the variance on the order of 2\%.  In practice we have averaged our results for $dS_P/da_\vk$ over eighty independent time slices.

\begin{figure}[t!]
\centerline{\scalebox{0.8}{\includegraphics{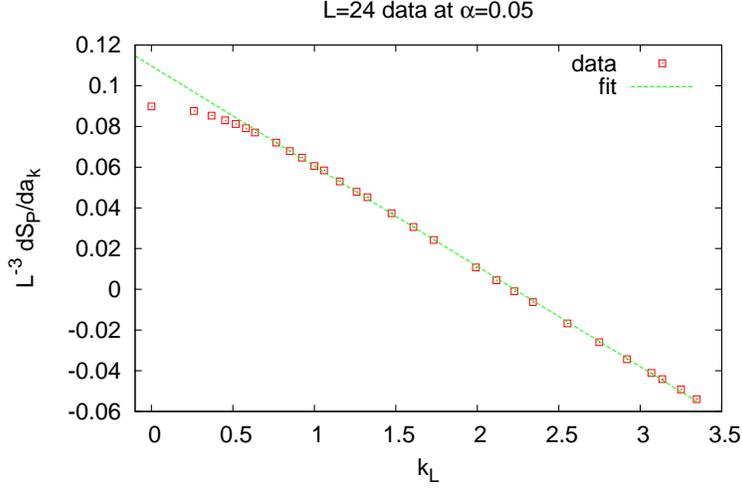}}}
\caption{Derivatives of the PLA $L^{-3} \pa S_P/\pa a_{\vk}$ evaluated at $a_{\vk}=\a=0.05$, vs.\ lattice momenta $k_L$.  Also shown is a linear best fit to the data at $k_L > 0.7$.}
\label{d24a05}
\end{figure} 

\begin{figure}[t!]
\centerline{\scalebox{0.8}{\includegraphics{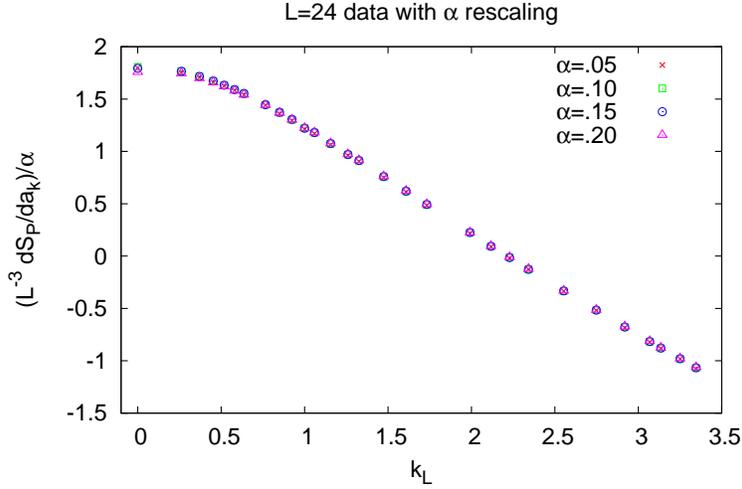}}}
\caption{Derivatives $L^{-3} (\pa S_P/\pa a_{\vk})_\a$ divided by $\a$, vs.\ lattice momenta $k_L$, 
for $\a=0.05,0.10,0.15,0.20$.  It is clear that the derivatives of $S_P$ depend linearly on $\a$.}
\label{scaling}
\end{figure} 

\begin{figure}[h!]
\centering
\subfigure[~ L=12]{
\resizebox{79mm}{!}{\includegraphics{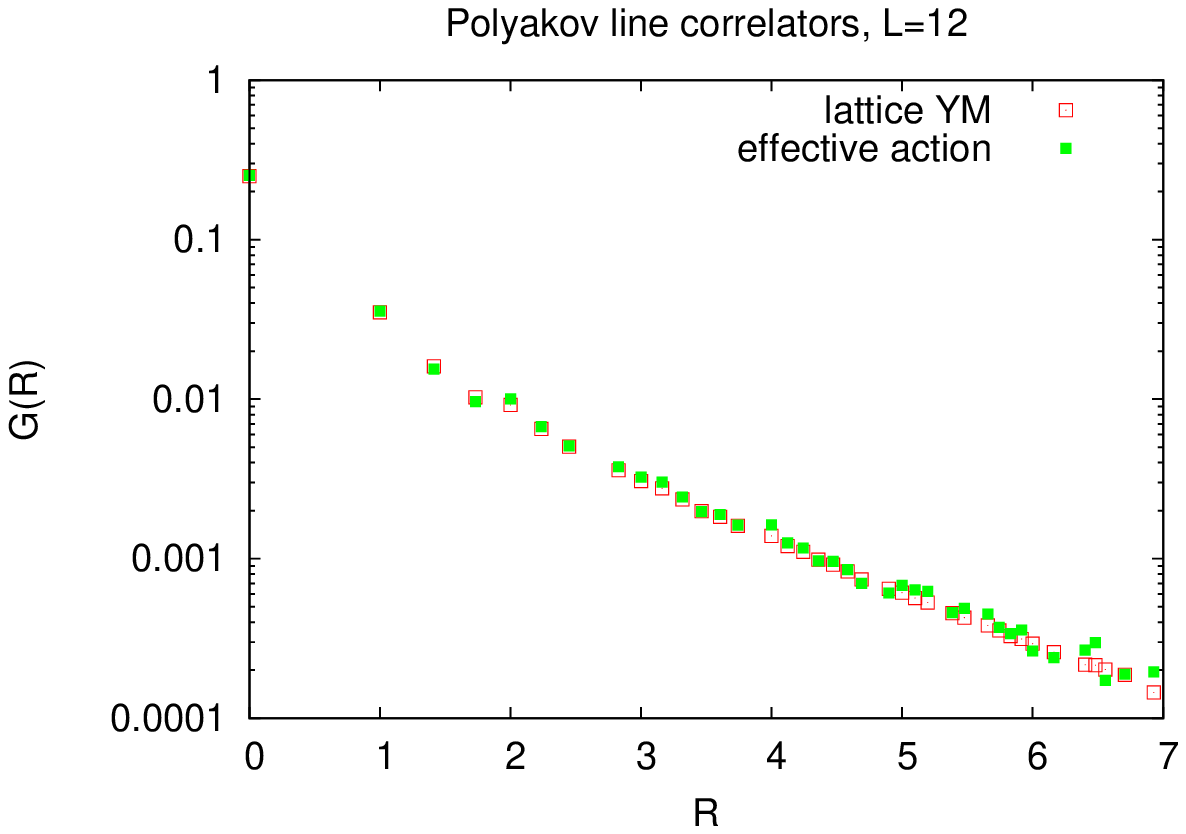}}
\label{corr12}
}
\subfigure[~ L=16]{
\resizebox{79mm}{!}{\includegraphics{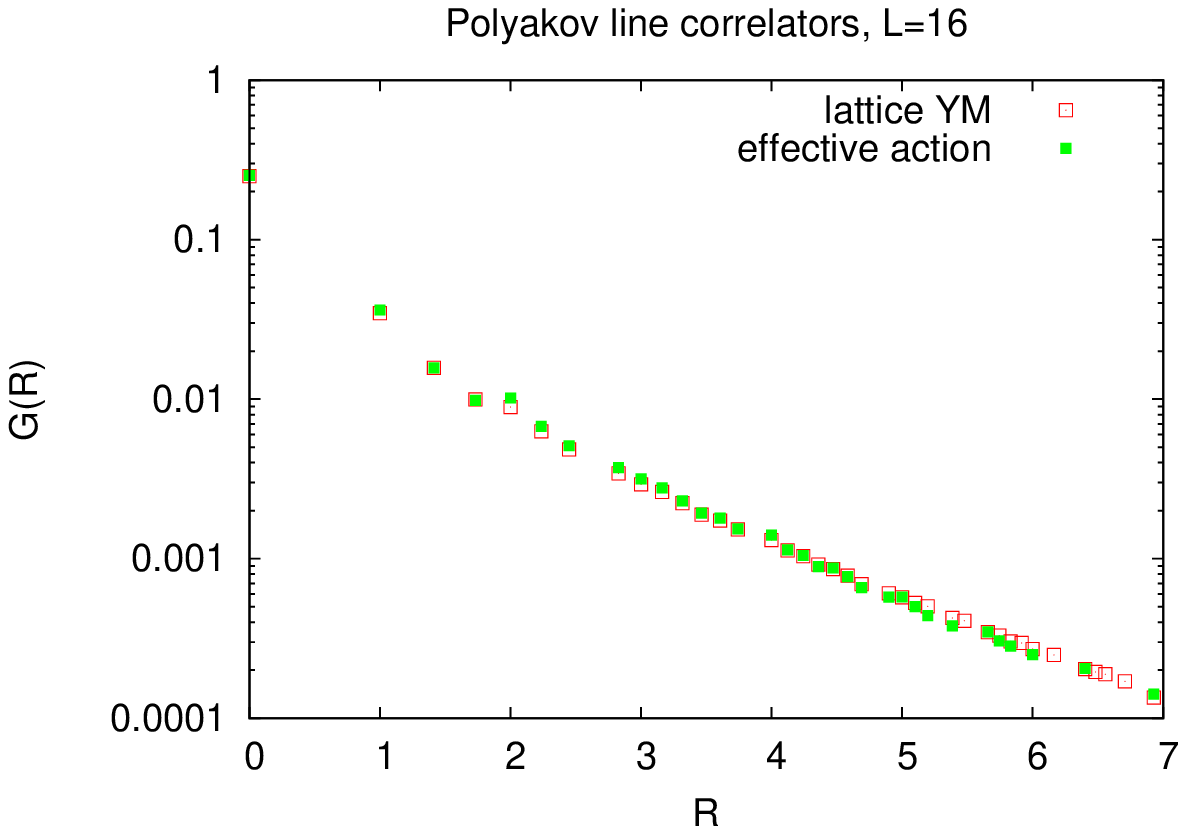}}
\label{corr16}
}
\subfigure[~ L=20]{
\resizebox{79mm}{!}{\includegraphics{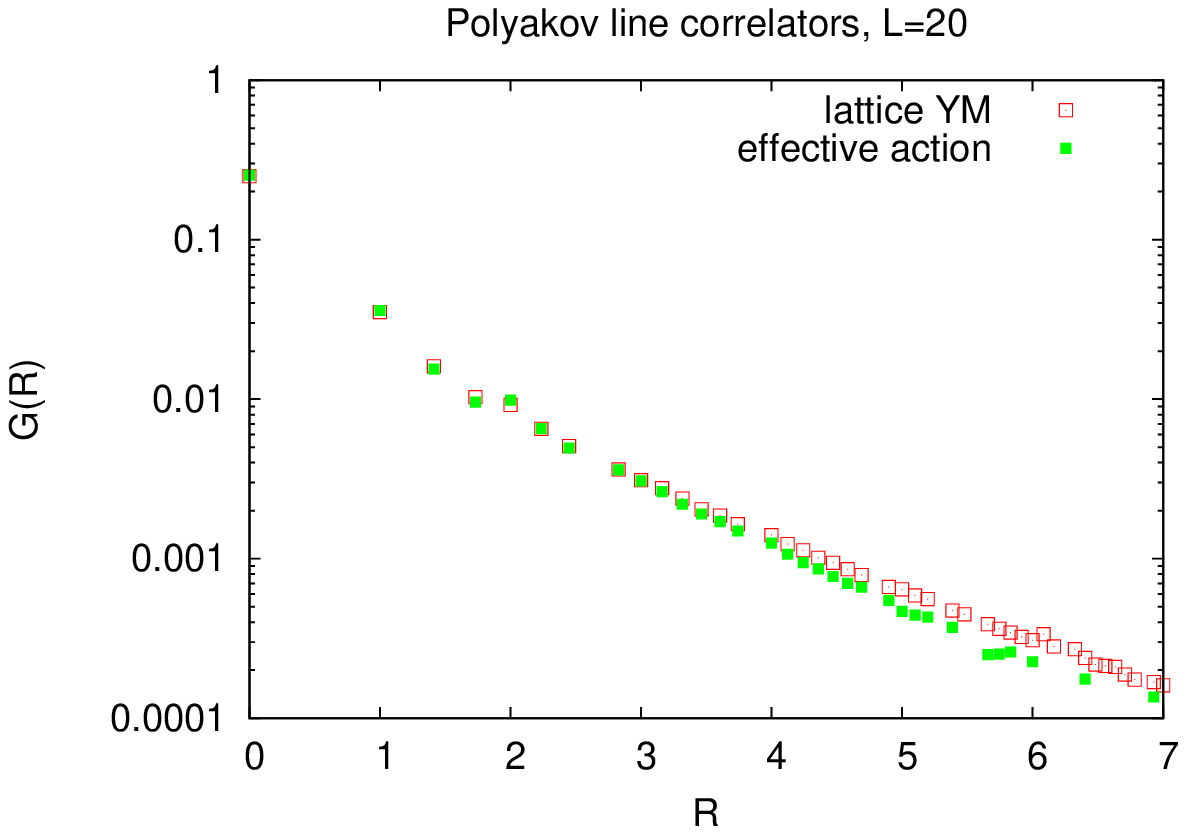}}
\label{corr20}
}
\subfigure[~ L=24]{
\resizebox{79mm}{!}{\includegraphics{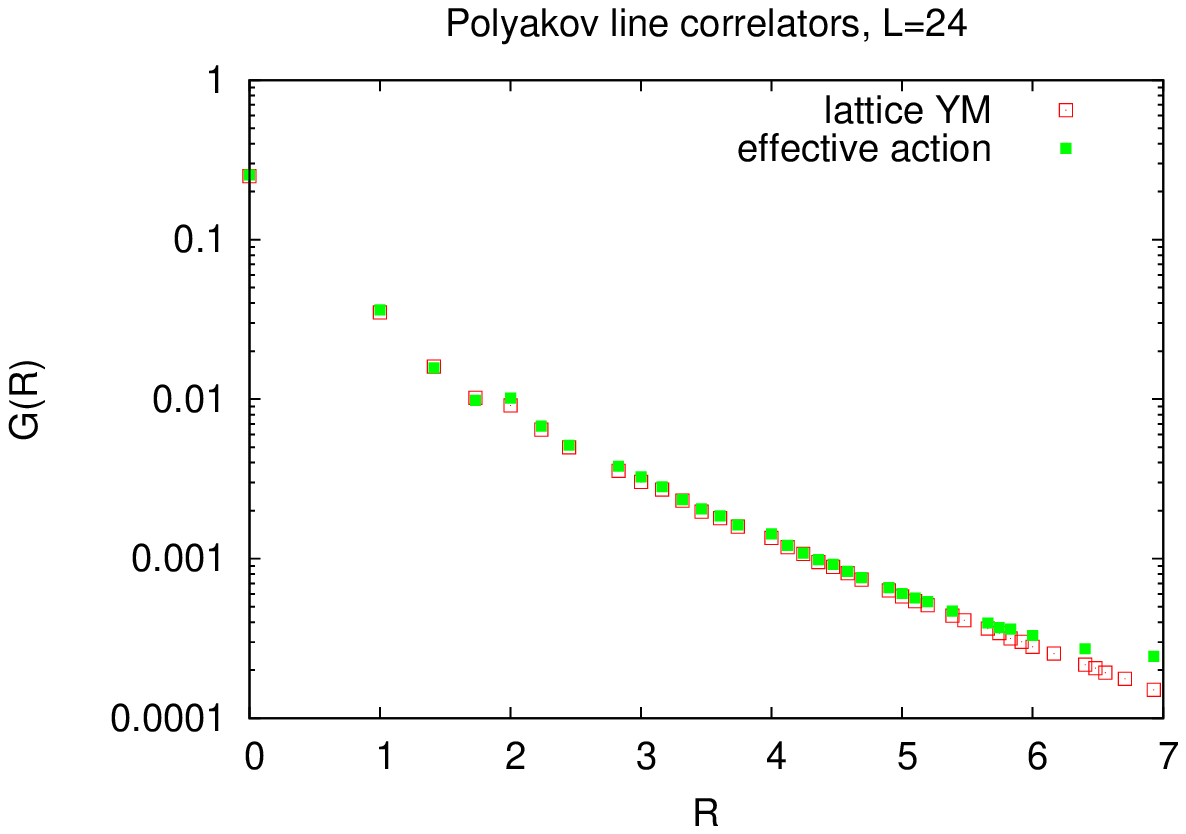}}
\label{corr24}
}
\caption{A comparison of the Polyakov line correlation functions $G(|\vx-\vy|) = \langle P_{\vx}  {P_\vy} \rangle$ as computed
via lattice Monte Carlo simulation of the underlying gauge theory on a $L^3 \times 4$ lattice at coupling $\b=2.2$, and via Monte Carlo simulation of the corresponding effective action $S_P$ of eq.\ \rf{SP1}.  Lattices are of spatial extension $L=12,16,20,24$
lattice spacings. Note that off-axis displacements are included.}
\label{Pcorr}
\end{figure}

    In Fig.\ \ref{d24a05} we display our results for the $L^{-3} dS_P/da_\vk$ vs.\ $k_L$ at $\a=0.05$, and lattice spatial extension $L=24$.  (Note that, apart from Figs.\ \ref{Pcorr} and \ref{cross}, all our figures show data derived at an $L=24$ extension.) The 
underlying SU(2) lattice gauge theory is defined as a Wilson action on a periodic $24^3 \times 4$ volume, at the coupling 
$\b=2.2$.  The calculations were made, in this case, at lattice momenta with components $k_i = 2\pi m_i/L$, with
the following $(m_1 m_2 m_3)$ triplets:
\bea
& &\Bigl\{ (000),(100),(110),(111),(200),(210),(211),(300),(311),(320),
\non \\
       & &         (400),(322),(421),(430), (333),(433),(443),(444),(554), (654), 
\non \\
 & &  (655), (665),  (766),  (777),  (887), (988), (998), (10\; 9 9), (10\; 10\; 10)\Bigr\} \ .
\eea
On this plot the data point displayed at $k_L=0$ is a factor of two smaller than the actual data value; this was done for reasons to be explained shortly.

The striking thing about this data is that, for lattice momenta $k_L > 0.7$, the data points clearly fall on a straight line.
The second fact is that the data is linearly proportional to $\a$ at these small $\a$ values, as we see in Fig.\ \ref{scaling}.  In this figure we divide  $dS_P/da_\vk$ by $\a$, at $\a=0.05,0.10,0.15,0.20$, and find that the data points coincide.  The linearity
of the derivative w.r.t.\ $a_\vk$ implies that the action itself is quadratic in these variables, leading to a simple
bilinear form 
\bea
           S_P =  \oh c_1 \sum_{\vx} P^2_{\vx} -  2c_2 \sum_{\vx \vy} P_{\vx} Q(\vx - \vy) P_{\vy} \ ,
\label{SP1}
\eea
where
\bea
Q(\vx-\vy)  &=& {1\over L^3} \sum_{\vk} \tQ(k_L) e^{i \vk \cdot (\vx - \vy)} \ .
\eea
This leads to derivatives
\bea
{1\over L^3} \left({d S_P[U_\vx(a_{\vk})] \over da_{\vk}}\right)_{a_{\vk}=\a} &=& \left\{ 
   \begin{array}{cc}
      \a(\oh c_1 - 2c_2 \tQ(k_L)) & k_L \ne 0 \cr
       & \cr
      2\a(\oh c_1 - 2c_2 \tQ(0))   &   k_L=0 \end{array} \right. \ .
\label{dS_P}
\eea       
The relative factor of two in the $k_L=0$ and $k_L > 0$ cases is due to the fact that $\sum_{\vx} 1 = L^3$, while
$\sum_{\vx} \cos^2(\vk \cdot \vx) = \oh L^3$.  The $k_L > 0$ data should extrapolate, as $k_L \ra 0$, to a value which
is half the result at $k_L=0$, which is why we have divided the derivative at $k_L=0$ by a factor of 2, when displaying
these values on Figs. \ref{d24a05} and \ref{scaling}.  The constants $c_1$ and $c_2$ are obtained from a linear fit to the
data at $k_L > 0.7$, as shown in Fig.\ \ref{d24a05}.\footnote{In practice we fit the data for each $\a$, at $k_L>0.7$, to the form
 $A(\a) - B(\a) k_L$.  We then fit  $A(\a),~B(\a)$ to straight lines, and the constants $c_1,~c_2$ are extracted from the
slopes, i.e.\ $dA/d\a = \oh c_1$, and $dB/d\a = 2 c_2$.  The choice of $0.7$ as the lower limit is a potential source of systematic error, since the value for $c_1$ can vary up to $1\%$ when the lower limit is increased (the variation of $c_2$ is smaller).
We find, however, that the choice of $0.7$ as the lower limit minimizes the reduced $\chi^2$ value of the linear fit.}

\begin{figure}[htb]
\centerline{\scalebox{0.7}{\includegraphics{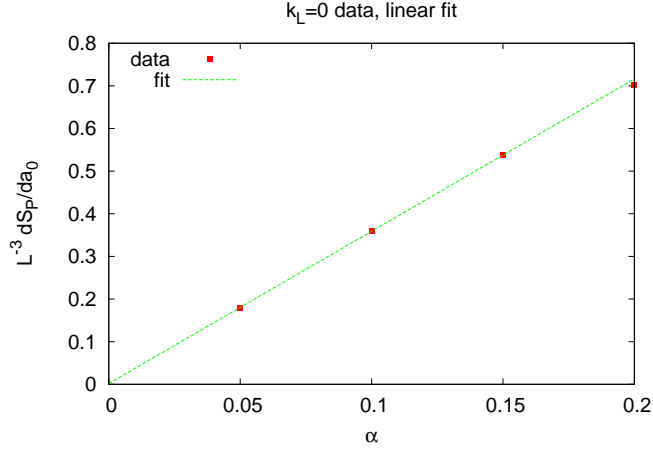}}}
\caption{The derivatives of $S_P$ with respect to the amplitude of the zero mode, evaluated at several values of $\a$.
The slope of this data is used to determine $r_{max}$ of the bilinear kernel $Q(\vx-\vy)$, as explained in the text.}
\label{zmode}
\end{figure} 

    It is clear that for $k_L > 0.7$, the $k$-space kernel is $\tQ(k_L)=k_L$.  If this were true at all $k_L$, then we would
have $Q=\sqrt{-\nabla_L^2}$ in position space, where $\nabla^2_L$ is the lattice Laplacian.  However,
a kernel of this kind is infinite range, which would violate one of the assumptions of the Svetitisky-Yaffe analysis (cf.\ 
\cite{Svetitsky:1985ye}).  In any case, $\tQ(k_L)$ deviates from linearity at small momentum.  We therefore make
an ansatz for the kernel which imposes the finite range restriction on $Q$ in a simple way:
\bea
           Q(\vx-\vy) = \left\{  \begin{array}{cc}
                     \Bigl(\sqrt{-\nabla_L^2}\Bigr)_{\vx \vy}  &  |\vx-\vy| \le r_{max} \cr
                       0 & |\vx-\vy| > r_{max} \end{array} \right. \ .
\label{Q}
\eea
Given $r_{max}$,  $\tQ(k_L)$ is obtained by a Fourier transform of $Q(\vx-\vy)$.  To determine $r_{max}$, we do a 
linear fit of the $k_L=0$ data
\bea
        {1\over L^3} \left({d S_P \over da_0}\right)_{a_0=\a} \ ,
\eea
as shown in Fig.\ \ref{zmode}.  Let the slope of the line be $D$.  Then, from \rf{dS_P},
\bea
              c_1 - 4c_2 \tQ(0)) = D \ ,
\eea
and we choose $r_{max}$ to satisfy this condition as closely as possible.  We then have $\tQ(k_L)$ at all $\vk$.  

\begin{figure}[htb]
\centerline{\scalebox{0.8}{\includegraphics{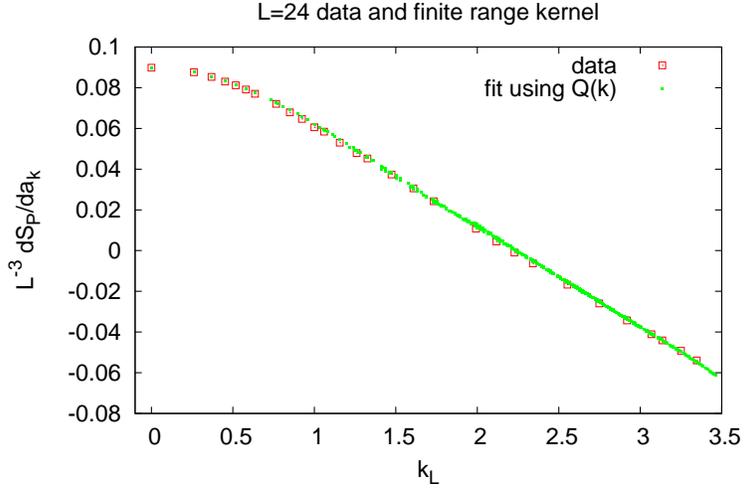}}}
\caption{A test of eq.\ \rf{dS_P} at $\a=0.05$.  The derivative data of Fig.\ \ref{d24a05} is plotted against
the conjectured fitting function $\a(\oh c_1 - 2c_2 \tQ(k_L))$ with $r_{max}=3$.}
\label{kernel}
\end{figure} 

    In Fig.\ \ref{kernel} we plot the data shown in Fig.\ \ref{d24a05}  together with the values computed for
\bea
             \a(\oh c_1 - 2c_2 \tQ(k_L))
\eea
(cf.\ eq.\ \rf{dS_P}) at $\a=0.05$.  Agreement seems to be quite good in the entire range of $k_L$.

    We have repeated this analysis at smaller volumes of spatial extension $L=12,16,20$.  The results for 
$c_1,c_2,r_{max}$ are shown in Table \ref{tab1}.  In Table \ref{tab2} we record non-zero components of $Q(\vx)=(-\nabla^2_L)_{\vx \mathbf{0}}$ up to $|\vx| < 3.2$ and lattice volume $24^3$.  For $r_{max}=3$, the last entry in the table should be replaced by $Q(3,1,0)=0$.  The rest of the non-zero elements of $Q$ are obtained from the table via permutation symmetry, $x_i \leftrightarrow x_j$, and reflection symmetry 
$x_i \ra -x_i$, among the coordinate components.

\begin{table}[h!]
\begin{center}
\begin{tabular}{|c|c|c|c|} \hline
         $ L $ &  $c_1$   &  $c_2$ & $r_{max}$ \\
\hline
        12 & 4.364(6) & 0.491(1) & 3.2  \\ 
        16 & 4.417(4) & 0.498(1) & 3.0 \\
        20 & 4.416(7) & 0.493(1) & 3.0 \\
        24 & 4.414(8) & 0.493(1) & 3.0 \\
        \hline
\end{tabular}
\caption{Constants defining the effective Polyakov line action for pure YM theory, $L^3 \times 4$ lattice, $\b=2.2$.} 
\label{tab1}
\end{center}
\end{table}

\begin{table}[h!]
\begin{center}
\begin{tabular}{|c|c|c|c|} \hline
         $x_1 $ &  $x_2$   &  $x_3$ & $Q(\vx)$ \\
\hline
        0 & 0 & 0 & 2.38760  \\ 
        1 & 0 & 0 & -0.22001 \\
        1 & 1 & 0 & -0.02357 \\
        1 & 1 & 1 & -0.00774 \\
        2 & 0 & 0 & -0.01279 \\
        2 & 1 & 0 & -0.00455 \\
        2 & 1 & 1 & -0.00246 \\
        2 & 2 & 0 & -0.00160 \\
        2 & 2 & 1 & -0.00111 \\
        3 & 0 & 0 & -0.00200 \\
        3 & 1 & 0 & -0.00121 \\
        \hline
\end{tabular}
\caption{Non-zero elements of the bilinear kernel $Q(\vx)$ at $r_{max}=3.2$ and $L=24$.} 
\label{tab2}
\end{center}
\end{table}

\subsection{Comparing the PLA to the underlying lattice gauge theory}    

    We now have a concrete proposal for the effective Polyakov line actions at various spatial volumes $L^3$ ranging from
$12^3$ to $24^3$, and which correspond to an underlying lattice SU(2) gauge theory at $\b=2.2$ on an $L^3 \times 4$ lattice volume. The actions are specified by eqs.\ \rf{SP1}, \rf{Q}, and the constants in Table I.  Above the lattice
volume $12^3$, these actions are about the same.

\begin{figure}[h!]
\centerline{\scalebox{0.5}{\includegraphics{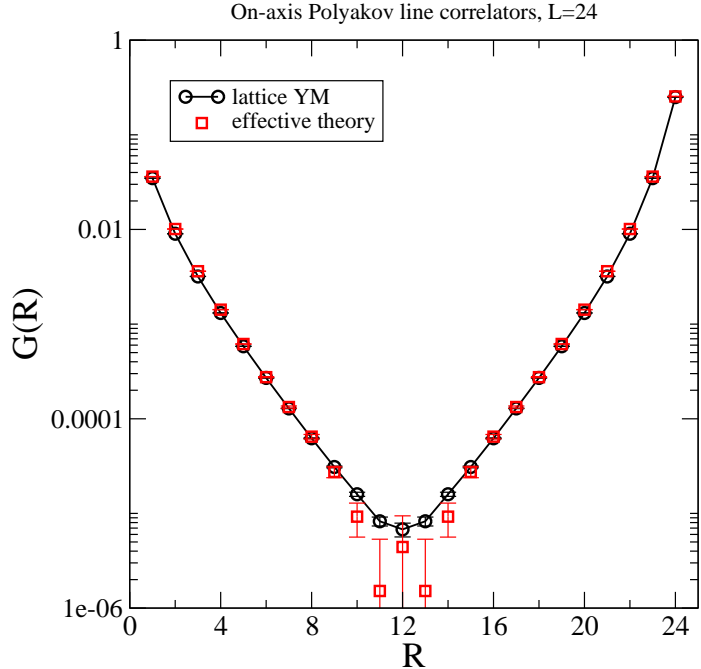}}}
\caption{A high-statistics comparison of the Polyakov line correlation function $G(|\vx-\vy|) = \langle P_{\vx}  {P_\vy} \rangle$ computed for the lattice gauge and effective theories, for displacements $\vx-\vy$ parallel to the $x,y$ or $z$-axes, and
spatial volume $24^3$.}
\label{corrax}
\end{figure} 

    The crucial question, of course, is whether these proposed Polyakov line actions are correct; they are certainly different from
the action suggested in ref.\ \cite{Greensite:2012dy}, which was derived for gauge configurations in a rather unrepresentative region of configuration space.  There is one obvious and essential test:  Do the derived Polyakov line actions reproduce the Polyakov line correlator calculated in the corresponding lattice gauge theory?  Thus
we compute, via numerical simulation of the Polyakov line action at $L=12,16,20,24$,
\bea
            G(|\vx-\vy|) = \langle P_{\vx}  {P_\vy} \rangle \ ,
\label{PP}
\eea
and compare the result to the same observable obtained from standard lattice Monte Carlo at $\b=2.2$ on an 
$L^3 \times 4$ volume.  The result for these four cases is shown in Fig.\ \ref{Pcorr}. 
Note that off-axis displacements are included, with $xyz$-components of the displacement $\vx-\vy$ in the range $[-4,4]$.

    The data in Fig.\ \ref{Pcorr} is limited to displacements of magnitude $R < 7$ lattice spacings, and it is interesting to
consider larger displacements for larger lattices.   This calls for higher statistics.  In Fig.\ \ref{corrax} we compare the
results for $G(\vx-\vy)$ obtained on a $24^3$ lattice for the effective theory, and on a $24^3 \times 4$ lattice for the
SU(2) gauge theory, again at $\b=2.2$.   In this figure we show the results for displacements $\vx-\vy$ parallel to any one of the coordinate axes.  The L\"uscher-Weisz noise reduction method \cite{Luscher:2001up} was used in obtaining the Polyakov line correlator in
the lattice gauge theory, while for the effective Polyakov line action the correlator was obtained from 38,400 configurations (about two orders of magnitude more than was used in Fig.\ \ref{Pcorr}).  
 
    The agreement of the correlators in the PLA and the underlying lattice gauge theory seen in Fig.\ \ref{corrax} is extraordinary,
and it persists down to magnitudes of order $10^{-5}$.\footnote{The Polyakov line correlator derived from the Inverse Monte Carlo method in ref.\ \cite{Dittmann:2003qt,*Heinzl:2005xv} was displayed on a linear, rather than logarithmic, scale, and hence the precision of agreement with lattice gauge theory, in that approach, is difficult to judge.}   While this is not a proof that $S_P$
is the correct effective action, it is difficult to believe that agreement of Polyakov line correlators to this level of precision is coincidental.

\section{\label{sec:gh}Polyakov line action for an SU(2) gauge-Higgs system}

\begin{figure}[h!]
\centerline{\scalebox{0.7}{\includegraphics{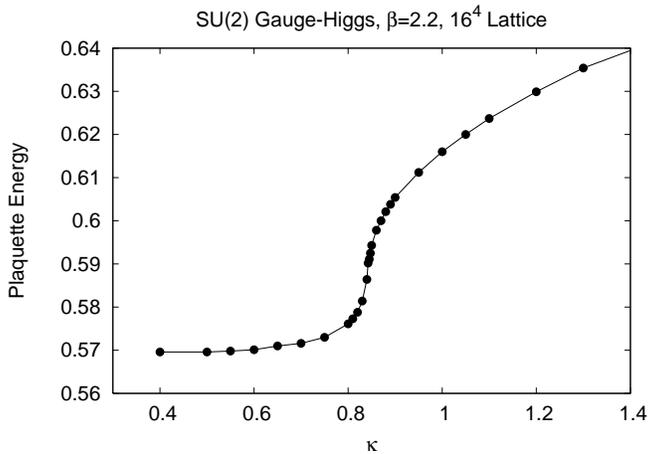}}}
\caption{Plaquette energy vs.\ gauge-Higgs coupling $\k$ at fixed $\b=2.2$, for the SU(2) gauge-Higgs theory with fixed Higgs modulus on a $16^4$ lattice volume, showing a sharp crossover at $\k \approx 0.84$.}  
\label{cross}
\end{figure}

     We now add a scalar matter field in the fundamental representation of the gauge group, thereby breaking
explicitly $Z_2$ center symmetry.  The simplest case is a fixed-modulus Higgs field, and for SU(2) gauge theory
the action can be written in the following way:
\bea
    S = \b \sum_{plaq} \oh \mbox{Tr}[UUU^{\dg}U^{\dg}] + \k \sum_{x,\m} \oh
              \mbox{Tr}[\phi^\dg(x) U_\m(x) \phi(x+\widehat{\m})] \ ,
\label{gh_action}
\eea
where $\phi(x)$ is SU(2) group-valued. The work of Fradkin and Shenker \cite{Fradkin:1978dv},
itself based on a theorem by Osterwalder and Seiler \cite{Osterwalder:1977pc}, demonstrated that the Higgs region and
the ``confinement-like" regions of the $\b-\k$ phase diagram are continuously connected.   Subsequent Monte Carlo studies found that there is only a single phase at zero temperature (there might have been a separate Coulomb phase), although there is a line of first-order transitions between the confinement-like and Higgs regions, which eventually turns into a line of sharp crossover around ${\b=2.775,\k=0.705}$, cf.\ \cite{Bonati:2009pf} and references therein.   
At $\b=2.2$ the crossover occurs at $\k \approx 0.84$, as seen in the plaquette energy data shown in Fig.\ \ref{cross}.  There is also a steep rise in the Polyakov line expectation value as $\k$ increases past this point.       

   We will work at $\b=2.2$ on a $24^3 \times 4$ lattice volume, but this time at Higgs coupling $\g=0.75$,
which places us in the ``confinement-like'' phase a little below the crossover point.  For these parameters, the Polyakov line has a VEV of $\langle P_{\vx} \rangle =0.0515$.  Once again, we generate sets of thermalized Polyakov line holonomies,
and compute $L^{-3} dS_P/da_{\vk}$ as explained in the previous section.  

\begin{figure}[ht]
\centering
\subfigure[~ full range]{
\resizebox{79mm}{!}{\includegraphics{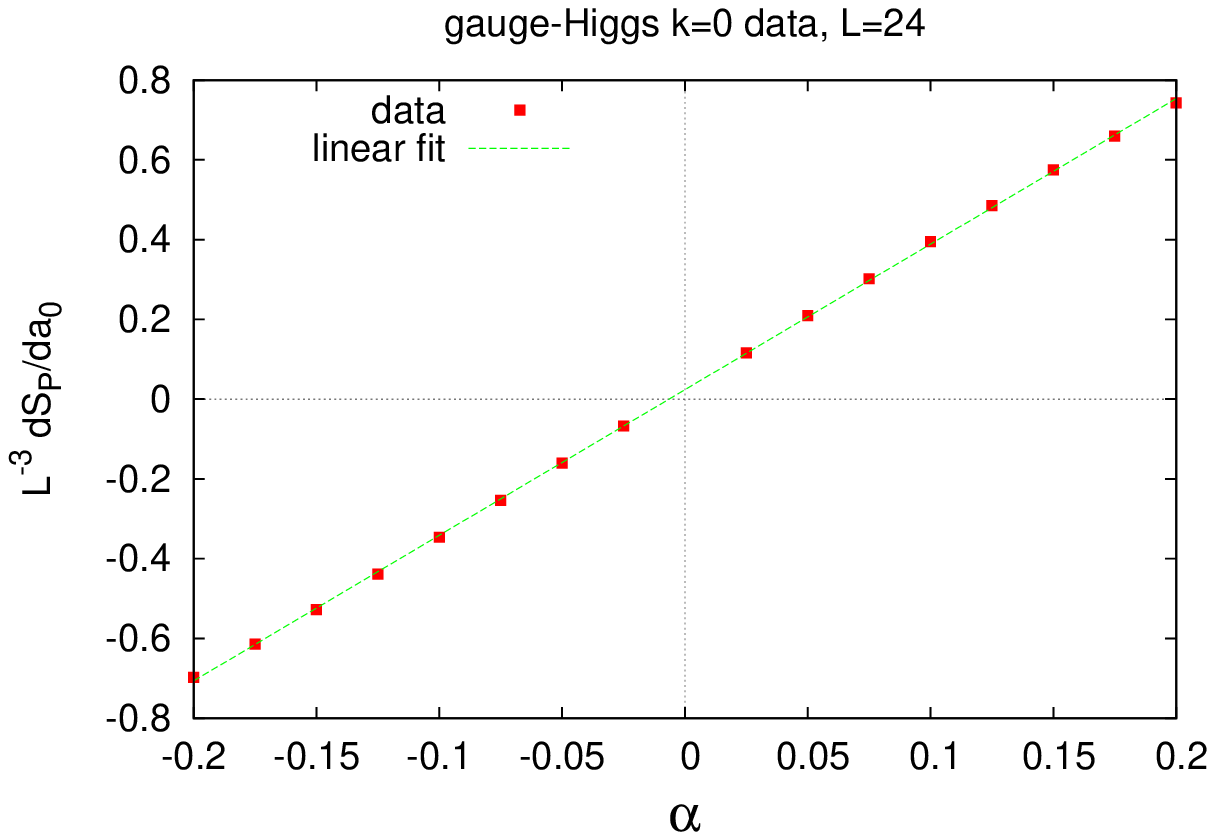}}
\label{zmodeh1}
}
\subfigure[~ close-up]{
\resizebox{79mm}{!}{\includegraphics{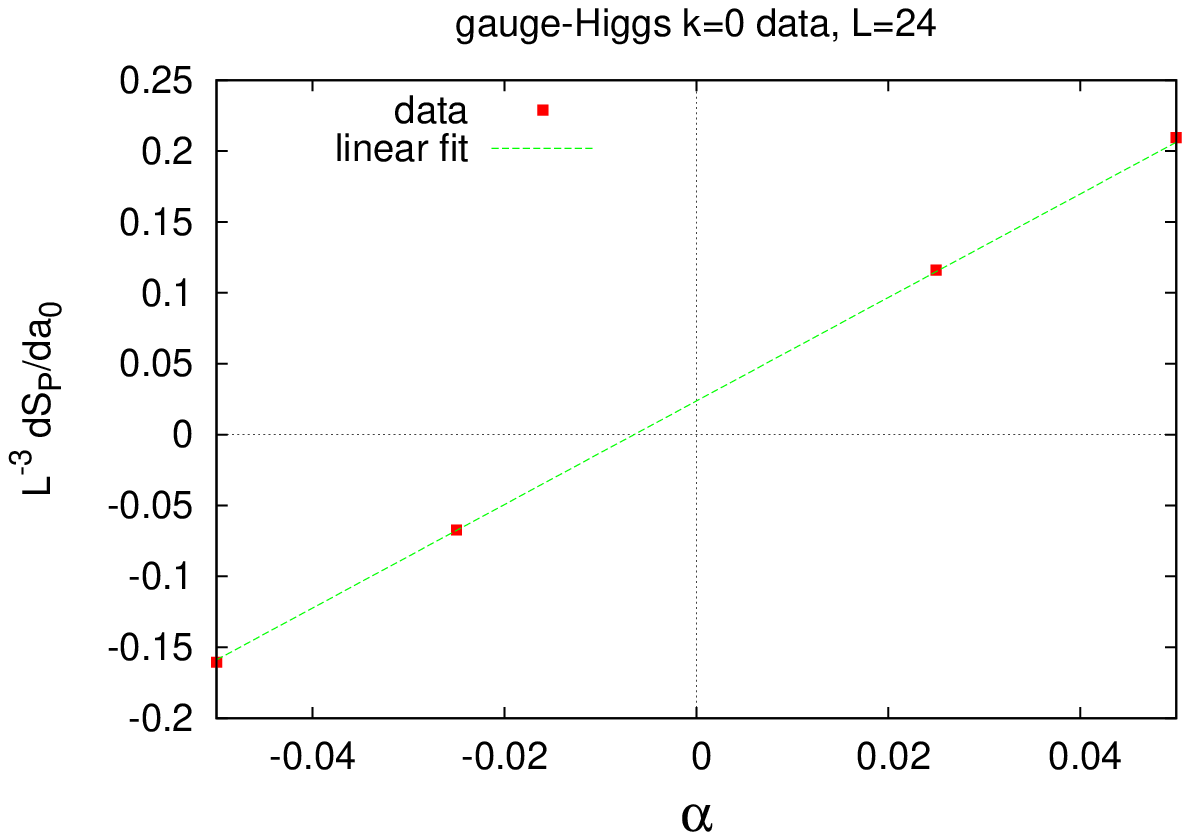}}
\label{zmodeh2}
}
\caption{The derivatives of $S_P$ with respect to the amplitude of the zero mode in the gauge-Higgs theory, evaluated at positive and negative values of $a_0=\a$.  (a) shows the full range of the data; (b) is a closeup near $\a=0$.  The $y$-intercept of this data is non-zero, and
determines the coefficient $c_0$ of the linear, $Z_2$-symmetry breaking term in the effective PLA \rf{SP2}.}
\label{zmodeh}
\end{figure}    

\begin{figure}[h!]
\centerline{\scalebox{0.8}{\includegraphics{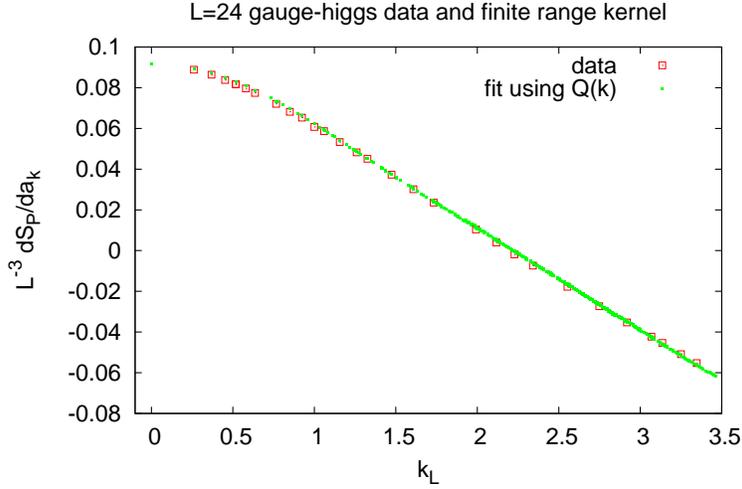}}}
\caption{Same as Fig.\ \ref{kernel} for the gauge-Higgs theory.  We plot the data for the derivative
$L^{-3} \pa S_P/\pa a_{\vk}$ vs. $k_L$ against
the conjectured fitting function $ \a(\oh c_1 - 2c_2 \tQ(k_L))$ with $r_{max}=3.2$. }
\label{kernelh}
\end{figure}

    The derivatives $\pa S_P/\pa a_{\vk}$ at $a_{\vk}=\a$ are computed as before, and at each $k_L > 0$ the results are simply 
proportional to $\a$.  The constants $c_1,c_2$ are again extracted by a linear fit to the $k_L > 0.7$ data.  However, the data at $k_L=0$ is not strictly proportional to $\a$; there is also an $\a$-independent constant contribution to the data.  This fact can be seen in Fig.\ \ref{zmodeh}.  The straight line is a best fit to $L^{-3} \pa S_P/\pa a_0$ evaluated at $a_0=\a$ for $\a$ in the range 
$[-0.2,0.2]$.  The $y$-intercept of this line does not pass through zero, but rather through $y=0.0236(14)$.  This implies that
$S_P$ must contain a term which is linear in $P_\vx$, i.e.
\bea
           S_P =   c_0 \sum_{\vx} P_{\vx} + \oh c_1 \sum_{\vx} P^2_{\vx} -  2c_2 \sum_{\vx \vy} P_{\vx} Q(\vx - \vy) P_{\vy} \ ,
\label{SP2}
\eea 
and it is clear from inspection that $c_0$ must equal to the $y$-intercept in Fig.\ \ref{zmodeh}.  It is also clear that only the
$k_L=0$ mode contributes to the linear term, and is therefore invisible in the derivatives of $S_P$ at $k_L>0$.   We define
$Q(\vx - \vy)$ again by \rf{Q}, with $r_{max}$ determined as in the pure-gauge theory.  The final set of parameters for the effective Polyakov line action is given in Table \ref{tab3}, and we plot the $k_L>0$ data, together with the quantity
\bea
             \a(\oh c_1 - 2c_2 \tQ(k_L))  ~~\text{vs.}~~~ k_L
\eea
in Fig.\ \ref{kernelh}.

\begin{table}[h!]
\begin{center}
\begin{tabular}{|c|c|c|c|c|} \hline
         $ L $ &  $c_0$   &  $c_1$ & $c_2$ & $r_{max}$ \\
\hline
        24 & .0236(14) & 4.447(9) & 0.501(1) & 3.2  \\ 
        \hline
\end{tabular}
\caption{Constants defining the effective Polyakov line action for gauge-Higgs theory, $24^3 \times 4$ lattice,
$\b=2.2,~\k=0.75$.}
\label{tab3} 
\end{center}
\end{table}
 
\begin{figure}[h!]
\centerline{\scalebox{0.4}{\includegraphics{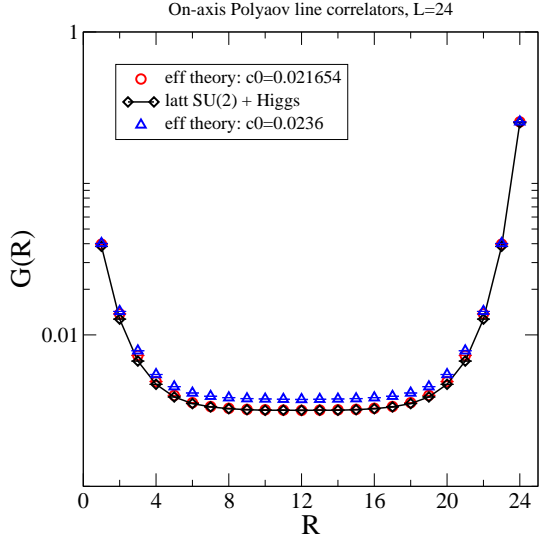}}}
\caption{A comparison of the Polyakov line correlation functions $G(|\vx-\vy|) = \langle P_{\vx}  {P_\vy} \rangle$ as computed
via lattice Monte Carlo simulation of the underlying gauge-Higgs theory (black diamonds) on a $24^3 \times 4$ lattice, 
at couplings $\b=2.2,~\k=0.75$, and via Monte Carlo simulation of the corresponding effective action $S_P$ of eq.\ \ref{SP2}
(blue triangles, $c_0=0.0236$).  Also shown is a simulation of the effective action with a slightly different value of $c_0=.02165$ (red circles).}
\label{corrH24}
\end{figure}  

    As in the pure gauge theory, the crucial test is to see whether the Polyakov line correlator \rf{PP} found from numerical
simulation of the gauge-Higgs theory \rf{gh_action} agrees with the same observable computed in the derived Polyakov line
action \rf{SP2}.   The results are shown in Fig.\ \ref{corrH24}. In this case the agreement between the lattice gauge Higgs
correlator (black diamonds) and the correlator of the effective action (blue triangles), while fairly close, is not perfect.  However,
the result for the effective action depends very sensitively on the value of $c_0$, and of course there is an errorbar associated
with this quantity.  In our best fits, $c_0=0.0236(14)$.  With a little trial and error, one can find a value of $c_0$ for the
effective effective action such that the corresponding correlator (red circles) agrees almost exactly with the gauge-Higgs value.
This happens at a value $c_0=0.02165$ which is not far outside our errorbars, about 1.4 $\s$ away from $c_0=0.0236$.

\section{\label{sec:conclude}Conclusions}

     Motivated by the well-known sign problem, we have applied the relative weights method to determine the effective Polyakov line action $S_P$ for both pure and gauge-Higgs lattice SU(2) gauge theory.  This effective action turns out to be a remarkably
simple expression, which is bilinear in the Polyakov line variables $P_{\vx}=\oh \tr[U_{\vx}]$:
\bea
           S_P &=&   c_0 \sum_{\vx} P_{\vx} + \oh c_1 \sum_{\vx} P^2_{\vx} -  2c_2 \sum_{\vx \vy} P_{\vx} Q(\vx - \vy) P_{\vy}
\non \\
           Q(\vx-\vy) &=& \left\{  \begin{array}{cc}
                     \Bigl(\sqrt{-\nabla_L^2}\Bigr)_{\vx \vy}  &  |\vx-\vy| \le r_{max} \cr
                       0 & |\vx-\vy| > r_{max} \end{array} \right.  \ ,
\label{SP3}
\eea
with $c_0=0$ in the pure-gauge theory, and non-zero in the gauge-Higgs theory.  
Our results so far have been obtained at lattice coupling 
$\b=2.2$, and $N_t=4$ lattice spacings in the time direction.  The effective action has been checked by computing Polyakov line correlators in both the effective theory and the underlying gauge theory, and we have found that these correlators
agree quite well with each other.  This is especially true in the pure gauge theory, where agreement persists down to correlator values of order $10^{-5}$.  These results, together with previous checks
in the case of strong coupling \cite{Greensite:2012dy}, inspire some confidence that the method works.  
The next immediate step will be to understand how the couplings of the effective theory evolve as a function of coupling $\b$ and
temperature $1/N_t$.

     In the longer term our interest is the sign problem, and to address that problem it will be necessary to derive the effective Polyakov line action corresponding to lattice SU(3) gauge fields coupled to matter at zero chemical potential.  In lattice gauge theory fixed to temporal gauge, the chemical potential $\m$ is introduced via the replacement
\bea
U_0(\vx,t=0) \ra e^{N_t \m} U_0(\vx,t=0) ~~~,~~~ U^\dg_0(\vx,t=0) \ra e^{-N_t \m} U^\dg_0(\vx,t=0)  \ .
\label{intro_mu}
\eea   
It is not hard to see that, to all orders in the strong coupling + hopping parameter expansion, the
effective PLA obtained at $\m \ne 0$ is related to the action at $\m=0$ by a simple substitution
\bea
         S^{\m\ne 0}_P[U_{\vx},U^\dg_{\vx}] = S^{\m=0}_P[U_{\vx} \ra e^{N_t \m} U_{\vx}, U^\dg_{\vx} \ra e^{-N_t \m} U^\dg_{\vx}] \ .
\label{replace}
\eea
We will assume that this identity holds in general.  The strategy is then to apply one or more of the methods 
\cite{Gattringer:2011gq,*Mercado:2012ue,Fromm:2011qi,Aarts:2011zn,Greensite:2012xv}, which were developed for solving Polyakov line actions with a chemical potential, to our derived
effective action.  If the sign problem is tractable in the effective Polyakov line theory, as suggested by the earlier work
cited above, then it may be possible to extract useful results regarding the QCD phase diagram.

\acknowledgments{J.G.\ would like to thank Kim Splittorff for helpful discussions.  J.G.'s research is supported in part by the
U.S.\ Department of Energy under Grant No.\ DE-FG03-92ER40711.  K.L.'s research is supported by STFC under the DiRAC framework. We are grateful for support from the HPCC Plymouth, where the numerical computations have been carried out.}

\appendix*

\section{\label{quarks}Dynamical Fermions}

    In this appendix we will just sketch how the relative weights algorithm can be applied in the case where there
are two mass-degenerate dynamical fermions coupled to the gauge field.  

    Taking $D$ as the Dirac operator, and $M\equiv D^\dg D$, the integration measure is
\bea
          e^{S_L} = \det M ~ e^{S_W} \ ,
\eea
where $S_W$ is the Wilson (or other improved pure-gauge) action.  Defining 
\bea
   M^{(m)}= M\Bigl[U_0(\vx,0)=U^{(m)}_\vx,U_k(x,t)\Bigr] \ ,
\eea
we have
\bea
          \exp[\D S_P^{(m+1)}] &=&  {Z_{m+1} \over Z_m}
\non \\
             &=&  \left\langle { \det M^{(m+1)}  \over \det M^{(m)} } \exp[\D S_W^{m+1}] \right\rangle_m \ ,
\label{B1}
\eea
where $\langle ... \rangle_m$ signifies, as before, the expectation value in a probability measure $\exp[S_L^{(m)}]/Z_m$.  The VEV can be evaluated via hybrid Monte Carlo, but in this case there is the question of how to evaluate
the ratio of determinants in \rf{B1}.  In the most straightforward approach, defining $\d M^{(m+1)} = M^{(m+1)} - M^{(m)}$
and writing 
\bea
\det  M^{(m+1)}  = \det\Bigl\{ M^{(m)} ( \mathbbm{1} +  [M^{(m)}]^{-1}\d M^{(m+1)})\Bigr\} \ ,
\eea
it is not hard to see that
\bea
{ \det M^{(m+1)}  \over \det M^{(m)} } &=& \exp\left[ \tr \log \Bigl\{\mathbbm{1} +  [M^{(m)}]^{-1}\d M^{(m+1)} \Bigr\} \right]
\non \\
  &\approx& \exp\left[ \tr \Bigl\{ [M^{(m)}]^{-1}\d M^{(m+1)} \Bigr\} \right] \ .
\eea
Therefore
\bea
\exp[\D S_P^{(m+1)}] &=&  \left\langle  \exp\left[ \tr \Bigl\{ [M^{(m)}]^{-1}\d M^{(m+1)} \Bigr\} \right] e^{\D S_W^{(m+1)} }
\right\rangle_m \ .
\label{B2}
\eea
It is best not to evaluate the trace  $\tr \{ [M^{(m)}]^{-1}\d M^{(m+1)} \}$ directly, because of the computational expense, but rather indirectly, by the method of ``noisy pseudofermions'' (c.f.\ section 8.4 in ref.\ \cite{Gattringer:2010zz}).

    A way of completely avoiding $M^{-1}$ in the observable is to introduce another set of pseudofermions $\varphi$, distinct from the pseudofermions used by hybrid Monte Carlo, so that we may write
\bea
{ \det M^{(m+1)}  \over \det M^{(m)} } &=&  {\int D\varphi   \exp[- \varphi^\dg M^{(m)} \varphi] \over
                                                                    \int D\varphi   \exp[ -\varphi^\dg M^{(m+1)} \varphi] } \ .
\eea
Defining
\bea
        \D A^{(m+1)} =  \varphi^\dg M^{(m+1)} \varphi -   \varphi^\dg M^{(m)} \varphi  \ ,
\eea
we have
\bea                                           
   { \det M^{(m+1)}  \over \det M^{(m)} }  &=& \left\langle \! \! \!\left\langle \exp[\D A^{(m+1)} ] \right\rangle \! \! \! 
   \right\rangle_{m+1}  \ ,
\eea 
where $\langle \!\! \langle ... \rangle \!\! \rangle_{m+1}$ refers to the expectation value in the measure proportional to
\bea
     \exp[- \varphi^\dg M^{(m+1)} \varphi]  \ .
\eea    
The final result for the relative weight is
\bea
\exp[\D S_P^{(m+1)}] &=& \bigg\langle   \left\langle \! \! \!\left\langle \exp[\D A^{(m+1)} ] \right\rangle \! \! \! \right\rangle_{m+1}
       \exp[\D S_W^{(m+1)}] \bigg\rangle_m \ .
\label{B3}
\eea
Again, the expectation value $\langle...\rangle_m$ would be computed via the usual hybrid Monte Carlo algorithm.  Eq.\ \rf{B3} is to be used, as before, to compute path derivatives, hopefully leading to an expression for the effective PLA at $\m=0$.  The PLA
with a finite chemical potential would then be obtained from the substitution \rf{replace}.

    Apart from computation cost, we believe that the addition of dynamical fermions does not pose any problems of principle
to the derivation of the effective Polyakov line action via the relative weights approach.

\bibliography{pline}

\end{document}